# The Heuristic Dynamic Programming Approach in Boost Converters

Sepehr Saadatmand, Pourya Shamsi, and Mehdi Ferdowsi
Department of Electrical and Computer Engineering
Missouri University of Science and Technology
sszgz@mst.edu, shamsip@mst.edu, ferdowsi@mst.edu

*Abstract*—In this study, a heuristic dynamic programming controller is proposed to control a boost converter. Conventional controllers such as proportional-integral-derivative (PID) or proportional-integral (PI) are designed based on the linearized small-signal model near the operating point. Therefore, the performance of the controller during the start-up, the load change, or the input voltage variation is not optimal since the system model changes by varying the operating point. The heuristic dynamic programming controller optimally controls the boost converter by following the approximate dynamic programming. The advantage of the HDP is that the neural network–based characteristic of the proposed controller enables boost converters to easily cope with large disturbances. An HDP with a well-trained critic and action networks can perform as an optimal controller for the boost converter. To compare the effectiveness of the traditional PI-based and the HDP boost converter, the simulation results are provided.

## I. INTRODUCTION

In the past few decades, power electronics DC–DC converters have matured into ubiquitous technologies. DC–DC power converters are used in a wide variety of applications, such as electronic devices like tablets and laptops, and in aerospace and power systems. The growth of renewable energy sources (RESs), such as uninterruptible power supplies (UPSs), wind turbines, and photovoltaics, has increased the interest on DC–DC power converters [1]-[3]. The climate-based characteristics of the renewable energy sources lead to output voltage disturbances when facing load variations. Therefore, there has been a greater variety of research studies on the control scheme of DC–DC power converters. The three most important categories of DC–DC power converters include (i) buck, (ii) boost, and (iii) buck–boost [4]-[7].

To connect these energy resources to the grid, DC–AC inverters are used. However, the voltage level provided by several energy sources, such as photovoltaics and fuel cells, is lower than the required voltage for the inverter; therefore, the voltage level needs to be increased by boost converters. Boost converters, also known as step-up converters, are basic DC–DC converters that convert energy from the primary side to the secondary side by increasing the output voltage [8], [9]. An intermediate unit is used to connect residential photovoltaics into the grid. For these reasons, boost converters have attracted a large variety of attention [10], [11].

Controlling power electronics converters is a challenging task due to their nonlinearity (hybrid) characteristics caused by the switching. In addition, specifically in boost converters with a right half-plane, stabilization is a concern. An undesired decrease in error bandwidth can overcome this drawback. Based on the control concept of boost converters, there are various categories as voltage control and current control, fixed frequency and unfixed frequency, linear or nonlinear controllers [12].

The most common approach to controlling a boost converter is based on tuning the pulse width modulation (PWM) that controls the switch position. Conventional proportional-integral (PI) or proportional-integral-derivative (PID) controllers are the most popular, thanks to their easy-to-implement characteristics. PI or PID-type controllers are designed based on the small-signal model of the averaging circuit. The small-signal model is the linearized model of the averaging circuit around a specific operating point. These types of controllers are designed for small perturbations, and their effectiveness is highly affected when facing a large signal disturbance. In other words, the performance of conventional-type controllers is not suitable when facing uncertainties or large disturbances [13].

The other popular controller for boost converters is known as sliding mode control (SMC), which was first introduced in [14]. The most highlighted feature of SMC is their inherent variable structure, and the most negative point is its variable switching frequency, which is a concern regarding electromagnetic interference (EMI) analysis [15]. Several studies have tackled the SMC approach to overcome its drawbacks and improve its performance. A PWM-based adaptive SMC is introduced in [16] that behaves like a traditional PWM controller with a fixed frequency; however, this method needs an auxiliary hysteresis block. An $H^\infty$ control is proposed in [17] to regulate a boost converter based on the sliding-mode current control. Although the SMC technique has several advantages, drawbacks such as EMI, chattering, and auxiliary blocks make it less compelling.

The enhancement in the state-of-the-art microcontroller and its affordability have increased the interest in nonlinear optimal controllers. Optimization techniques have been used in several power electronic applications [18]-[25]. Dynamic Programming (DP) and model predictive controller (MPC) have been implemented in different control applications. The first one derives an optimal law based on the Bellman's equation to optimize the cost-to-go function, and the latter minimizes the cumulative cost in a specific time horizon. Several studies have implemented MPC approaches [24], [25], but DP optimizers are hard to design and implement. Therefore, by the knowledge of the author, there have been no studies in implementation of DP DC–DC power converters. Approximate/adaptive dynamic Programming (ADP) tackles the drawback of DP by using artificial neural networks (ANNs) to solve the optimization problem. The adaptive critic design (ACD) is one of the subcategories of ADPs. The basic form of ACDs is known as heuristic dynamic Programming (HDP). HDPs are utilized to optimally control a system. ACD methods are used in power-frequency regulation of grid-connected virtual inertia-based inverters [24], [25].

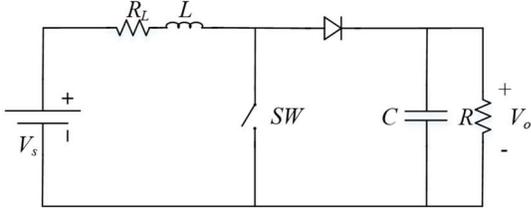

Figure 1. The circuit diagram of a simple boost converter

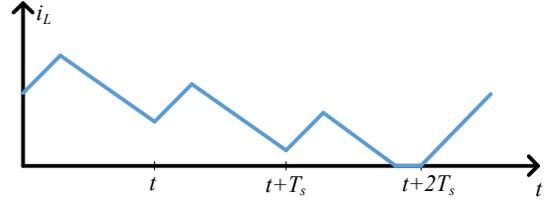

Figure 2. The inductor current mode describes the inverter mode: the converter performs in CCM mode in $t \in [t, t+T_s]$, and it operates in DCM mode in $t \in [t+T_s, t+2T_s]$

The highlighted contribution of this study is to propose a heuristic dynamic programming approach for the voltage regulation of a boost converter. The rest of the paper is organized as follows. Section II discusses the mathematical model of the boost converter. The heuristic dynamic programming, the training process, and implementation are explained in Section III. The simulation results are provided in Section IV to evaluate the performance and the effectiveness of the proposed controller. Lastly, the conclusion is presented in Section V.

## II. BOOST CONVERTERS

The circuit framework of a boost converter is illustrated in Figure 1. In this figure, *R, L,* and *C* are the load resistor, the input inductor, and the output capacitor, respectively. Two power electronics switches are used: a controllable switch, *Sw*, and a diode, *D*. The output voltage, which is typically fixed, is shown by $v_o$, and the input voltage, which is typically variable, is shown by $v_s$. The internal resistor of the inductor is also shown by $R_L$. In this model, the diode on-time resistance, equivalent series resistance of the capacitor, and switch on-time resistance are ignored. The state-space model of the system in a continuous-time region is presented. The discontinuous-time state-space model can be easily derived from the continuous-time model. The small-signal averaging model is not discussed in this section because the proposed controller is designed based on nonlinear systems.

The independent state vector that represents the proposed boost converter includes two variables: (i) the current of the inductor and (ii) the capacitor voltage [16], which can be defined as

$$x(t) = [i_L(t) \ v_o(t)]^T. \qquad (1)$$

Using the linear affine (linear plus offset), the proposed boost converter can be described by

$$\frac{dx(t)}{dt} = \begin{cases} A_1 x(t) + B v_s(t), & S = 1 \\ A_2 x(t) + B v_s(t), & S = 0, \text{ and } i_L(t) > 0 \\ A_3 x(t), & S = 0, \text{ and } i_L(t) = 0 \end{cases} \qquad (2)$$

where the state matrices can be defined as

$$A_1 = \begin{bmatrix} -\frac{R_L}{L} & 0 \\ 0 & -\frac{1}{RC} \end{bmatrix} \quad A_2 = \begin{bmatrix} -\frac{R_L}{L} & -\frac{1}{L} \\ \frac{1}{C} & -\frac{1}{RC} \end{bmatrix}$$

$$A_3 = \begin{bmatrix} -\frac{R_L}{L} & 0 \\ 0 & -\frac{1}{RC} \end{bmatrix} \quad B = \begin{bmatrix} \frac{1}{L} & 0 \end{bmatrix}^T. \qquad (3)$$

There are two main categories regarding the operating point in boost converters: (i) continuous conduction mode (CCM) and (ii) discontinuous conduction mode (DCM). In CCM mode, the inductor current is always positive regardless of the switch position, but in DCM mode, the current of the inductor is zero for a period of time when the switch is off. Figure 2 depicts a situation when the boost converter can perform in both CCM and DCM.

## III. HEURISTIC DYNAMIC PROGRAMMING

Optimization techniques have been used in a great variety of power electronics applications [26]-[29]. The conventional PI or PID controller for power electronics converters is restricted based on the following features. The parameters of the PI or PID controller require to be set online regarding the designer experience or the system response. Moreover, traditional controllers such as proportional controllers or integral controllers are proposed to perform in linear systems, and consequently in nonlinear systems their performances can be nonoptimal. The linearized model is used due to In order to design the parameters. Therefore, by variation in the operating point, the PI parameters are no longer optimal because the linearized model around the operating point is not valid. Lastly, the traditional PID or PI controllers are introduced for single-input single-output (SISO) systems; nonetheless, to control both the inductor current and the output voltage, they are not well suited for traditional controllers.

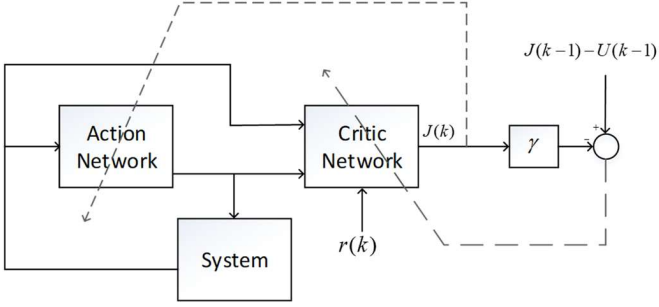

Figure. 3. Heuristic dynamic Programming block diagram

One of the most famous neural network–based controllers (neuro controls) is the adaptive critic design (ACD). ACDs are utilized to optimally control a system over time under conditions of noise and uncertainty. Inspired by the concept of reinforcement learning and approximate dynamic programming, a new category of optimization algorithm was introduced in [17].

Most significant ACDs can be itemized as one of the following: dual heuristic Programming (DHP), heuristic dynamic Programming (HDP), global dual heuristic Programming (GDHP), and global heuristic dynamic Programming (GHDP). Two main subnetworks are the core design of ACDs listed as: the critic and the action network. The critic network characteristics introduce the different types of ACDs.

The most straightforward form of ADCs can be presented as HDP. Figure 3 shows the design scheme of HDP. The estimation of the value (cost-to-go) function is the main goal of critic networks. In dynamic programming this function is known as the Bellman's equation. The goal of the action networks is to produce a set of control signals and to feed them to the system. The control signal and the state vector provide the critic network. As demonstrated in Figure 3, the dashed lines provide the required signals to train both the critic and the action network.

### A. Critic network

Inspired by the Bellman's equation, the value function can be written as

$$J(k) = \sum_{k=0}^{\infty} \gamma^i U(k+1) \tag{4}$$

to ensure that the cost-to-go is bounded where a discount factor is utilized as $\gamma$ $(0 < \gamma < 1)$. The utility function is represented by $U$, which in this paper is introduced as

$$U(k) = \sqrt{K_v e_v^2 + K_i e_i^2} \tag{5}$$

where $e_v$ and $e_i$ are the error signal for output voltage and the error of the current signal, respectively, that are defined as

$$e_v = v_{set} - v_o \tag{6}$$
$$e_i = i_{set} - i_L$$

To describe the significance of each error signal, $K_v$ and $K_i$ are defined as the output voltage coefficient and the inductor current coefficient, respectively. Another form of giving weight to the error signal parameters is through a normalized function with a weight matrix. A fully connected multi-layer forward neural network is utilized to perform as the critic network, as shown in Figure 4. The proposed neural network consists two hidden layers, and each hidden layer include five neuron. The critic network input is a vector including the output voltage, the inductor current, the output voltage error, the inductor current error, and the duty. The critic network input can be written as

$$IN_{\{critic\ network\}} = [v\ i_L\ e_v\ e_i\ d\ ]. \tag{7}$$

The proposed controller performs in real-time. Therefore, to optimize the cumulative error signal, the critic neural network is tuned forward in time. As mentioned earlier, using the Bellman's equation by minimizing the error between the utility function and two sequential value functions, the critic network is trained to estimate the cost-to-go function. Therefore, equation (8) needs to be minimized.

$$\sum_k [J(k) - \gamma J(k+1) - U(k)]^2. \tag{8}$$

In order to update the network weights the gradient decent technique can be utilized as follows

$$W_{critic}(k+1) = W_{critic}(k) + \Delta W_{critic}$$
$$\Delta W_{critic} = \alpha_c [J(k) - \gamma J(k+1) - U(k)] \frac{\partial J(k)}{\partial W_{critic}} \tag{9}$$

Where $\alpha_c$ and $W_{critic}$ are the learning rate and the critic network weight, respectively.

### B. Action network

In order to optimize the value function of the immediate future, the action network needs to generate the optimal set of control, it poses that the sum of the utility function needs to be minimized. A fully-connected feedforward neural network is utilized to implement the action network, analogous to the critic network. The action network input only includes the state variables and the state variables errors, which can be defined as:

$$IN_{\{action\ network\}} = [v\ i_L\ e_v\ e_i] \tag{10}$$

The duty cycle of the PWM is the action network output. The backpropagation algorithm is used to update the action network weights, in other words, the following expression needs to be minimized

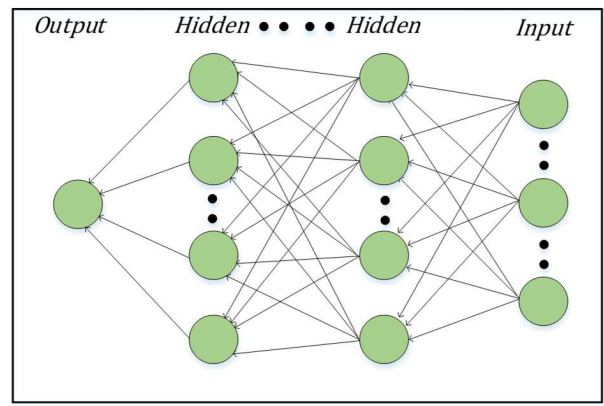

Figure. 4. Fully connected feedforward neural network

$$\zeta = \sum_k \frac{\partial J(k+1)}{\partial E(k)} \quad (11)$$

where $\alpha_a$ and $W_{action}$ present the learning rate and the action network weight, respectively. Therefore, the weights can be calculated as

$$\Delta W_{action} = -\alpha_a \zeta \frac{\partial \zeta}{\partial W_{action}} \quad (12)$$

$$W_{action}(k+1) = W_{action}(k) + \Delta W_{action}.$$

## IV. SIMULATION RESULTS

Recently, the fuel cell generation systems have attracted a great variety of attention because of their exclusive advantages such as high efficiency, no moving part, environment friendly, greater durability, and sustainability. Varying output voltage during the load changes can cause complicated control concerns. Therefore, a steady boost converter is essential that utilizes the fuel cell energy more efficiently and satisfies the requirement of a cascaded DC-AC inverter application. To evaluate the proposed controller, an HDP-based controller is implemented to regulate a boost converter. Figure 5 depicts the structure of the introduced HDP-based boost converter. As illustrated in this figure, both PI and HDP are implemented. The HDP signal is disabled when the critic neural network is pretrained. In other words, the state signal goes to the PI controller, and the PI controller regulates the output voltage. After utilizing the boost converter with random references of output voltage and load current, the training data (including the state and the duty cycle at each time step) is generated. After the critic network is pretrained, the HDP-based controller goes online and controls the boost converter. The critic network and the action network are updated at each control cycle. This control scheme includes both offline (to pretrain the critic network) and online learning (the online training process of both critic and action networks).

The parameters of the boost converter are shown in Table I. The performance of the boost converter at the start-up, the load change, and the change of the input voltage is evaluated, and a comparison between HDP and a PI controller is shown.

### A. Start-up

First, the dynamic behavior of the system during start-up under HDP and PI are analyzed. The start-up is under the nominal load (i.e., $P_{out}$ = 500 W, $R$= 80 Ω). Figure 6 illustrates the output voltage and the inductor current of the proposed boost converter during start-up for the HDP and PI controller, respectively. As expected, the system does not operate in its nominal operating point during transient time. As shown, the HDP controller performs much quicker, and the settling time regarding the HDP is $t_{set} \approx 5$ msec, but the settling time regarding the PI controller is greater than $t_{set} \approx 20$ msec. The voltage overshoot regarding the HDP controller (3%) is much less than that of PI controller (18%).

### B. Load change

To evaluate the performance of the proposed controller, a step-up load change scenario from 80 Ω to 200 Ω is simulated.

Table I. Boost converter parameters and information

| Parameter | Symbol | Value |
|---|---|---|
| Input voltage | $V_s$ | 60 ± 10% V |
| Output voltage | $V_o$ | 200 V |
| Output power | $P_{out}$ | 500 ±60% W |
| Load resistor | $R$ | 50 –200 Ω |
| Switching frequency | $f_{sw}$ | 20 kHz |
| inductor | $L$ | 860 µH |
| capacitor | $C$ | 860 µF |

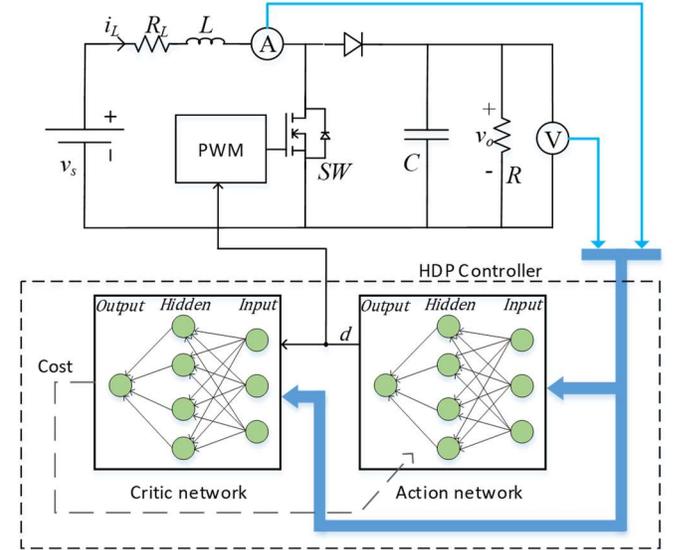

Figure 5. The block diagram of a HDP-based boost converter

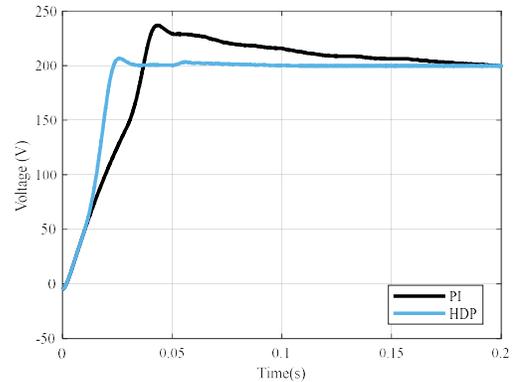

(a)

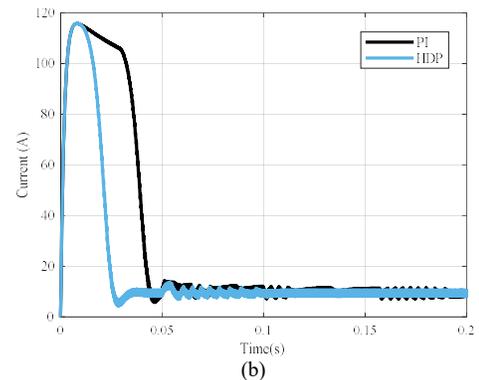

(b)

Figure 6. (a) the output voltage regarding the PI and HDP controller, (b) the inductor current regarding the PI and HDP controller

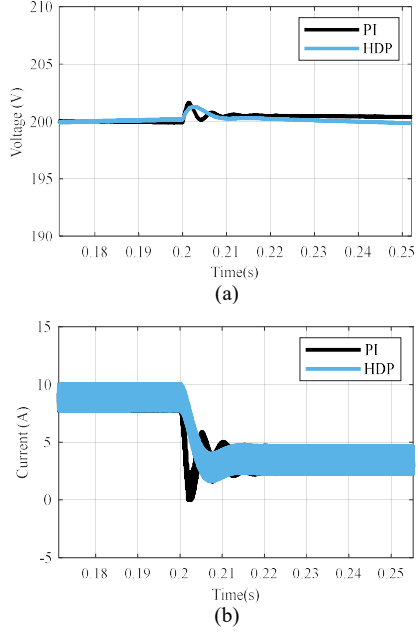

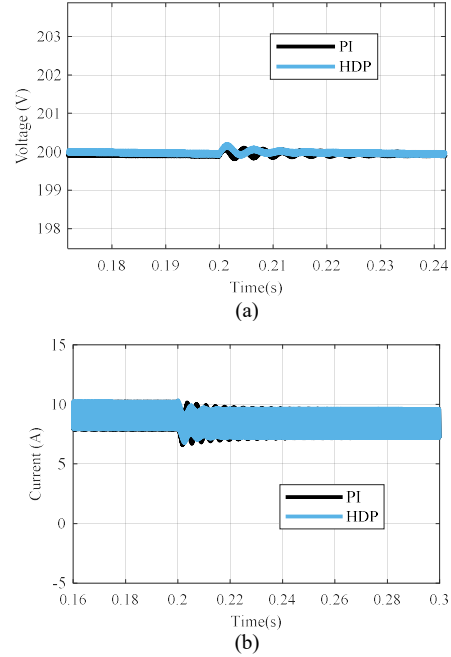

Figure 7. The performance of the boost converter in the load change, (a) the output voltage, (b) the inductor current

Figure 8. The performance of the boost converter in the reference voltage change, (a) the output voltage in reference voltage, (b) the inductor current

As previous simulations show, the PI controller does not function well when the performance of the boost converter is not near the nominal operating point. Figure 7 illustrates the output voltage and the inductor current of the boost converter under both the HDP and the PI controller, respectively. As shown, the HDP controller keeps regulating the voltage optimally, but the stable PI controller starts oscillating after the change in operating point.

*C. Input voltage change*

To evaluate the performance of the proposed controller regarding the input voltage changes, the maximum of reference voltage change is applied. The input voltage drops from 60 V to 54 V. Changing the reference voltage alters the linearized state-space model based on which the PI controller is designed. Therefore, the performance of the PI controller is not optimal. However, the HDP tracks the voltage reference with the minimum cumulative error at the optimal time horizon. Figure 8 depicts the voltage and the current output for both scenarios and for the HDP and the PI controller, respectively.

V. CONCLUSION

In this paper, a heuristic dynamic programming (HDP) approach was introduced to optimally control a boost converter. The model-free and neural network–based characteristics of the HDP algorithm enable the controller to perform with more robustness when facing large disturbances. The drawbacks of the conventional PI/PID controllers have been discussed facing large disturbances. A well-trained HDP controller can regulate the output voltage of a boost converter. The performance of the proposed HDP-based and PI controller is compared via simulations. The HDP controller exhibits a voltage regulation with more robustness and faster dynamics compared to traditional PI-based boost converters. By validating the effectiveness of the proposed controller in three different scenarios (i.e., during the start-up, load change, and input voltage variation), the proposed controller is introduced as a state-of-the-art control technique for boost converters.